\documentclass[amsmath,amssymb,nofootinbib]{revtex4}
\usepackage{graphicx}
\usepackage{epstopdf}
\usepackage{amsfonts}
\usepackage{amsmath}
\usepackage{amssymb}
\usepackage{subfigure}
\usepackage{hyperref}
\usepackage{url}
\usepackage{xcolor}
\usepackage{color}
\usepackage{mathrsfs}  
\usepackage{calrsfs}
\usepackage[font={scriptsize}]{caption}
\def\sideremark#1{\ifvmode\leavevmode\fi\vadjust{\vbox to0pt{\vss% the remark
 \hbox to 0pt{\hskip\hsize\hskip1em%                          will appear only
 \vbox{\hsize2cm\tiny\raggedright\pretolerance10000%          on the side
 \noindent #1\hfill}\hss}\vbox to8pt{\vfil}\vss}}}%
\definecolor{amaranth}{rgb}{0.9, 0.17, 0.31}
\definecolor{purple(munsell)}{rgb}{0.62, 0.0, 0.77}
\definecolor{americanrose}{rgb}{1.0, 0.01, 0.24}
\definecolor{palatinateblue}{rgb}{0.15, 0.23, 0.89}
\definecolor{royalblue(web)}{rgb}{0.25, 0.41, 0.88}
\definecolor{hanpurple}{rgb}{0.32, 0.09, 0.98}
\definecolor{beaublue}{rgb}{0.74, 0.83, 0.9}
\definecolor{carminered}{rgb}{1.0, 0.0, 0.22}
\definecolor{emerald}{rgb}{0.31, 0.78, 0.47}
\definecolor{vividviolet}{rgb}{0.62, 0.0, 1.0}
\definecolor{brightpink}{rgb}{1.0, 0.0, 0.5}
\hypersetup{ linktoc=all,
    colorlinks, linkcolor={palatinateblue},
    citecolor={brightpink}, urlcolor={amaranth}
}
\newcommand{\changeurlcolor}[1]{\hypersetup{urlcolor=#1}}    
\renewcommand{\d}[1]{\ensuremath{\operatorname{d}\!{#1}}}

\begin{document}

\title{Twisted Black Hole Is Taub-NUT}

\author{Yen Chin Ong}
     \email{ongyenchin@sjtu.edu.cn}
   
     \affiliation{%
Center of Astronomy and Astrophysics, Department of Physics and Astronomy,\\
Shanghai Jiao Tong University, Shanghai 200240, China.
}%

\begin{abstract}
Recently a purportedly novel solution of the vacuum Einstein field equations was discovered: it supposedly describes an asymptotically flat twisted black hole in 4-dimensions whose exterior spacetime rotates in a peculiar manner -- the frame dragging in the northern hemisphere is opposite from that of the southern hemisphere, which results in a globally vanishing angular momentum. Furthermore it was shown that the spacetime has no curvature singularity. We show that the geometry of this black hole spacetime is nevertheless not free of pathological features. In particular, it harbors a rather drastic conical singularity along the axis of rotation. In addition, there exist closed timelike curves due to the fact that the constant $r$ and constant $t$ surfaces are not globally Riemannian. In fact, none of these are that surprising since the solution is just the Taub-NUT geometry. As such, despite the original claim that the twisted black hole might have observational consequences, it cannot be. \newline \newline

%\noindent\fbox{\fbox{%
 %\parbox{0.76\textwidth}{%
     %   {\color{red}{\textbf{Remark:} In version 1 of this manuscript, we only discussed the various peculiar features of the twisted spacetime, and explained why it does not describe a smooth regular black hole. It has since been pointed out that this spacetime is none other than the Taub-NUT solution, expressed in a slightly different coordinate system. In this version, we have kept much of the discussion in version 1 since it may be of some interest to the readers, but we have modified and updated the manuscript with the mentions that the twisted black hole is in fact Taub-NUT spacetime.}}
    %}%
%}}

\end{abstract}

\maketitle
\section{Twisted Black Holes: What Happens to the Singularity Theorem?}

Recently, Zhang \cite{zhang} showed that general relativity admits the following vacuum solution in 4-dimensions (see also a follow-up work \cite{1610.00886}): 
\begin{flalign}\label{twisted}
g_{\text{twist}}=&-\frac{r^2-a^2-2Mr}{r^2+a^2}\d t^2 + \frac{4a(r^2-a^2-2Mr)\cos\theta}{r^2+a^2} \d t \d\phi + \frac{r^2 + a^2}{r^2-2Mr-a^2}\d r^2 + (r^2+a^2)\d \theta^2 \\ & + \frac{4a^4 + 8a^2Mr -4a^2r^2 - (3a^4 + 8a^2Mr-6a^2r^2 -r^4)\sin^2\theta}{r^2+a^2} \d \phi^2.
\end{flalign}
This should be compared to the well-known rotating solution, namely the Kerr metric:
\begin{flalign}
g_{\text{Kerr}}=&-\frac{r^2+a^2\cos^2\theta-2Mr}{r^2+a^2\cos^2\theta}\d t^2 - \frac{4aMr \sin^2\theta}{r^2+a^2\cos^2\theta} \d t \d\phi + \frac{r^2 + a^2\cos^2\theta}{r^2-2Mr+a^2}\d r^2 + (r^2+a^2\cos^2\theta)\d \theta^2 \\ & + \frac{(r^2+a^2)(r^2+a^2\cos^2\theta) + 2Mra^2\sin^2\theta}{r^2+a^2\cos^2\theta}\sin^2\theta \d \phi^2.
\end{flalign}\newline

In the Kerr metric, $a$ denotes the rotation parameter $a:=J/M$ where $J$ is the (ADM) angular momentum of the black hole. (We use the usual convention that the speed of light is set to unity: $c=1$.) In the twisted metric, $a$ is just a constant; it cannot be interpreted as a rotation parameter in the same manner because $J$ is identically zero (see below). 
One obvious difference is that some occurrences of $a^2\cos^2\theta$ in the Kerr metric have been replaced simply with $a^2$ in Zhang's twisted metric. In particular, the metric coefficient $g_{\theta\theta}$ is now independent of $\theta$. Furthermore, unlike the Kerr black hole event horizon, with is situated at $r_h=M+\sqrt{M^2-a^2}$, the twisted solution has an event horizon at $r_h=M+\sqrt{M^2+a^2}$ instead. Since the latter expression is always real, there is no upper bound on how large the parameter $a$ can be. This is not a problem since, as we mentioned, $a$ is not a rotation parameter in the usual sense. Also unlike the Kerr black hole which has complicated structures: an event horizon, an inner (Cauchy) horizon, and an ergosphere, the twisted black hole appears to lack all of these but the event horizon. In addition to the event horizon $r_h$, the denominator of $g_{rr}$ has another zero $r_- = M -\sqrt{M^2+a^2}$, which is negative for any non-zero real $a$. In the case of Kerr black hole, analytic continuation to negative $r$ is not considered physically realistic. See, e.g., the comment on p.13 in \cite{visser}. The geometry here is quite different from that of the Kerr solution, but for simplicity let us not consider $r<0$ any further in this work. (See \cite{1610.06135} for further discussion.)\newline

It was shown in \cite{zhang} that there is no frame dragging on the horizon, i.e., the horizon itself is static. However, the \emph{exterior} spacetime does experience frame-dragging (except on the equatorial plane), and the effect is maximum at some finite distance away from the black hole. The upper half spacetime above the equatorial plane rotates in the opposite direction than that of the lower half. Consequently, the ADM angular momentum $J$ \emph{vanishes} identically, though the Komar angular momentum does not \cite{zhang}. \newline
 
Unlike the Kerr black hole, the twisted black hole remains a black hole solution even if the ADM mass, $M$, is set to zero. Whether $M=0$ or not, the Kretschmann scalar $R^{abcd}R_{abcd}$ can be computed explicitly and shown to be finite everywhere \cite{zhang}. It is, surprisingly, independent of the angles $\theta$ and $\phi$:
\begin{flalign}\label{KS}
R^{abcd}R_{abcd}[g_{\text{twist}}] = &\frac{48 }{(r^2+a^2)^6}(a^8 -a^6M^2+12a^6Mr+15a^4M^2r^2-40a^4Mr^3-15a^2M^2r^4\\ \notag &+12a^2Mr^5  -15a^6r^2+15a^4r^4-a^2r^6+M^2r^6),
\end{flalign}
in contrast to the Kretschmann scalar of the Kerr black hole \cite{9912320}:
\begin{equation}
R^{abcd}R_{abcd}[g_{\text{Kerr}}] = \frac{48M^2}{(r^2 + a^2\cos^2\theta)^6}\left(r^6-15a^2r^4\cos^2\theta + 15a^4r^2\cos^4\theta - a^6\cos^6\theta\right).
\end{equation}
As mentioned in \cite{zhang}, it can been seen from Eq.(\ref{KS}) that the twisted black hole is regular everywhere, in the sense that there is no curvature singularity. This is unlike other vacuum black hole solutions in general relativity that harbors curvature singularity (either timelike or spacelike). \newline

The very \emph{absence} of a curvature singularity, however, raises a question: how does the gravitational collapse that resulted in the twisted black hole evade the singularity theorem of Penrose \cite{penrose}? (See \cite{jose} for a recent review). Essentially, Penrose proved that if a spacetime contains a non-compact Cauchy hypersurface $\Sigma$ (``causality condition'') and a closed future-trapped surface $\mathcal{H}$ (``trapping condition''), and if the null Ricci condition $R_{ab}U^aU^b \geqslant 0$ holds for all null vector $U$, then there exist future incomplete null geodesics. That is to say, light rays hit a singularity in finite affine time.
(The null Ricci condition is better known by the name ``null energy condition'', but it is really a geometric condition. See \cite{jose}, and also \cite{parikh}.) Note that a vacuum solution such as the twisted black hole satisfies the null Ricci condition trivially because $R_{ab} = 0$. \newline

Of course, strictly speaking, the Penrose singularity theorem is about singularities in the sense of geodesic completeness; it does not have anything to say about curvature singularities. However, in the case of black holes in general relativity, these two notions of singularity usually coincide (see also Section 5.1.5 of \cite{jose} and the references therein). In fact, if there exists incomplete timelike or null geodesics, we should ask where and why does the geodesic end \emph{physically}. For example, a Minkowski spacetime with one point artificially removed is not geodesically complete, but this is not very physical. On the contrary, it is quite natural that geodesics end when something very drastic happens to the geometry, such as when curvature blows up. Therefore, the fact that the twisted black hole has no curvature singularity at least suggests that the geometry should also be geodesically complete. This is therefore in tension with the singularity theorem, \emph{unless} the geometry of this black hole spacetime contains other pathologies, such as a conical singularity.  (A famous example of a spacetime that contains a conical singularity is that of a cosmic string. However, a cosmic string induces a conical defect due to its energy density. The twisted black hole is quite different as it is a vacuum solution.)  Indeed, 
an important premise in the singularity theorem is that the metric has to be ``sufficiently nice''. Technically this means that the metric has to be twice differentiable, i.e., of class $C^2$. An extension to metric of class $C^{1,1}$, i.e., the first derivative is Lipschitz continuous, has been achieved recently \cite{1502.00287}.  Despite the discussion in this paragraph, which motivated our investigation, we will see later on that the twisted spacetime \emph{is indeed geodesically incomplete} despite it not satisfying the conditions for the singularity theorem to apply -- this geodesic incompleteness arises from the spacetime pathology itself.\newline

We will show that the twisted black hole harbors some kind of conical singularity, which means that the metric tensor is not differentiable everywhere, which would explain why the singularity theorem does not apply. Furthermore, we will show that the exterior geometry of the twisted black hole is rather peculiar: the 2-dimensional surfaces of constant $r$ and constant $t$ are not globally Riemannian, with metric signature changing from $(+,+)$ to $(-,+)$ sufficiently near the poles. This change of signature is due to the change of sign in the metric coefficient $g_{\phi\phi}$. This indicates that the twisted spacetime harbors closed timelike curves (CTCs), which is yet another reason why the singularity theorem does not apply (the ``causality condition'' is not satisfied). \newline

If the twisted black hole solution is in fact new to general relativity, then it is a remarkable discovery and we should indeed try to find observational evidences for such a black hole. However, the presence of CTC means that this is probably not a physical solution that one might observe in the real universe. As it turned out, \emph{the solution is in fact not new either}. If one computes the Petrov type of the twisted black hole, one finds that it is of Petrov type D. However, \emph{all} asymptotically flat vacuum solutions of Petrov type D are already known \cite{kinnersley} -- there are only 10 of them -- and therefore the twisted black hole \emph{cannot} be a new solution. We show below that it is just the Taub-NUT (Taub-Newman-Unti-Tambourino) solution \cite{nut1,nut2,nut3}, which is known to exhibit the aforementioned pathological properties, and belongs to Petrov type D. \newline

As such, the claim of \cite{zhang} that the twisted black hole might have observational consequences is very unlikely.
Note that the Taub-NUT solution is indeed known to be geodesically incomplete (see, e.g., \cite{MKG, 1002.4342}. However, see \cite{1508.07622}), although it does not have a curvature singularity. 
\newline

\section{The Peculiar Geometry of Twisted Black Holes}

It is easier to focus our discussion on the massless case. As we shall see later, the inclusion of the mass term does not qualitatively affect our conclusions.\newline

For $M=0$, the event horizon is located at $r_h=a$ \cite{zhang}.
Let us consider constant $t$ and constant $r$ surfaces. For the massless black hole, we thus obtain the following 2-dimensional geometry:
\begin{equation}\label{2geo}
\d s^2 = (r^2+a^2)\d\theta^2 + \frac{4a^4-4a^2r^2 - (3a^4 -6a^2r^2 -r^4)\sin^2\theta}{r^2 + a^2} \d\phi^2.
\end{equation}
Let us call this surface $S_{a,r}$. The indices emphasize that for a fixed $a$, we have a family of surfaces indexed by $r$. \newline

We start by examining the Gaussian curvature of this 2-geometry. As we shall see, 
the Gaussian curvature across the horizon at the poles is not continuous. The expression for the Gaussian curvature, $K$, of $S_{a,r}$ is rather complicated in general and is not particularly illuminating to include here. At the poles, and assuming $a=1$, the expression simplifies to
\begin{equation}
K=\frac{4r^6+20r^4-36r^2+12}{16r^6-16r^4-16r^2+16}.
\end{equation}\newline

\begin{figure}[h!]
\captionsetup{justification=raggedright}
\begin{center}
\includegraphics[width=4in]{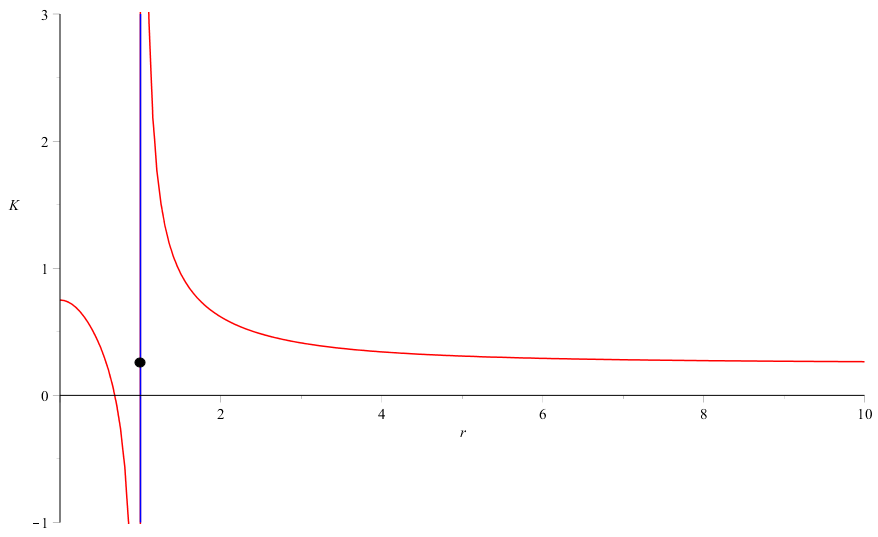} 
\caption{The plot of the Gaussian curvature at the poles for the family of surfaces $S_{a,r}$, here with $a=1$.  Outside the horizon, $r>a$, the Gaussian curvature is positive and approaches infinity as it tends toward the horizon. Inside the horizon, $r<a$, the Gaussian curvature is positive near the origin $r=0$, but becomes negative and eventually tends toward negative infinity as it tends toward the horizon. The Gaussian curvature on the horizon ($r=a$; denoted by the vertical line) is finite and is equal to $1/4a^2$, here denoted by a black dot. \label{gaussian}
} 
\end{center}
\end{figure}

We provide a plot in Fig.(\ref{gaussian}). On the horizon itself, i.e., when $r=a$, the Gaussian curvature is $1/4a^2$, this is just the Gaussian curvature of a 2-sphere of radius $2a$. This is as expected because the 2-geometry defined by the metric \ref{2geo} reduces to precisely such a sphere when $r=a$. 
Outside the horizon we find that the Gaussian curvature is positive, and as one approaches the horizon, it diverges: $K \to +\infty$. Inside the horizon, $K$ is initially positive near $r=0$, but becomes negative, and eventually tends to $-\infty$ as one approaches the horizon. This shows that $K$ is not continuous at the horizon in a very extreme manner, despite being finite on the horizon. This in turn implies that $S_{a,r}$ does not have a ``nice'' geometry at the poles, when $r\neq a$.\newline

Let us now examine in details the 2-geometry in two regions: the interior spacetime and the exterior spacetime, i.e., inside and outside the event horizon, respectively.\newline

%%%%%%%%%%%%%%%%%%%%%%%%%%%%%%%%%%%%%%%%%%%%%%%%%
\subsection{Interior Geometry}

A conical singularity at some point $p \in S_{a,r}$ means that the total angle going around $p$ is not equal to $2\pi$. If it is less than $2\pi$, we say that the conical singularity has an angle deficit, whereas if it is more than $2\pi$, then we say that the conical singularity has an angle excess.  \newline

Let us take $p$ to be the north pole of this geometry (by symmetry everything we discuss below also applies to the south pole). Consider a circle centered at $p$. Then we can compute the quantity ``circumference/radius'', which we shall denoted by $\Psi$:
\begin{equation}\label{Psi}
\Psi=\frac{2\pi\sqrt{g_{\phi\phi}}}{\displaystyle \int_0^\theta{\sqrt{g_{\theta'\theta'}}} \d\theta'}
=
\frac{2\pi\sqrt{4a^4-4a^2r^2-(3a^4-6a^2r^2-r^4)\sin(\theta)^2}}{\theta(a^2+r^2)}.
\end{equation}
In the absence of any conical singularity at the pole, this quantity should reduce to $2\pi$ in the limit $\theta \to 0$.
However, one finds that 
\begin{equation}
\Psi = 4\pi a\frac{\sqrt{a^2-r^2}}{(a^2+r^2)\theta}  -\frac{1}{2}\frac{\pi \theta (3a^4-6a^2r^2-r^4)}{(a^2+r^2)\sqrt{a^2-r^2}} + O(\theta^3).
\end{equation}
At the pole, $\theta = 0$ and only the first term remains. However, as is evident from the expression, the first term is not well-behaved at the pole:
\begin{itemize}
\item[(1)] Inside the horizon, $r < a$, the term is real and divergent.
\item[(2)] Outside the horizon, $r > a$, the term is imaginary and divergent.
\end{itemize}
In fact the only place where $\Psi$ is well-defined at the pole, is on the horizon itself: there Eq.(\ref{Psi}) reduces to $\Psi = 2\pi \sin\theta/\theta$, which, in the limit $\theta \to 0$, is exactly $2\pi$. This is, again, as expected, since the 2-geometry defined by the metric \ref{2geo} reduces to a 2-sphere of radius $2a$ precisely when $r=a$. \newline

\begin{figure}[!h]
\captionsetup{justification=raggedright
}
\centering
\mbox{\subfigure{\includegraphics[width=3in]{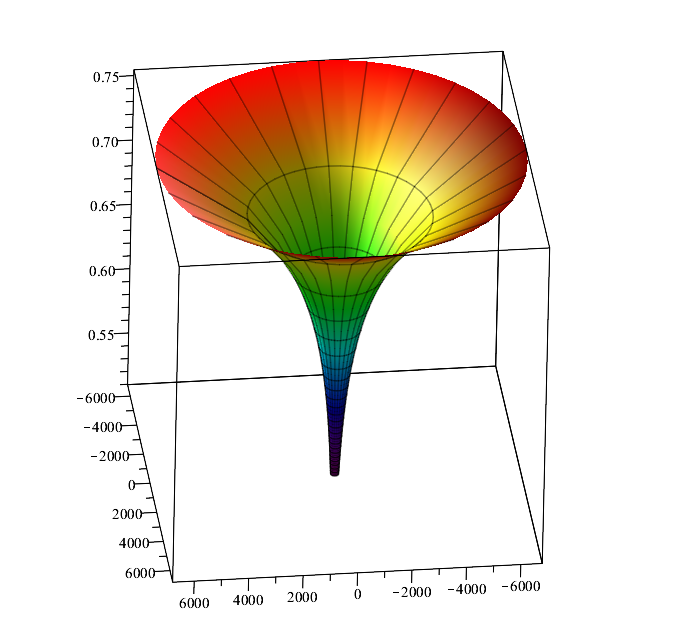}}\quad
\subfigure{\includegraphics[width=3in]{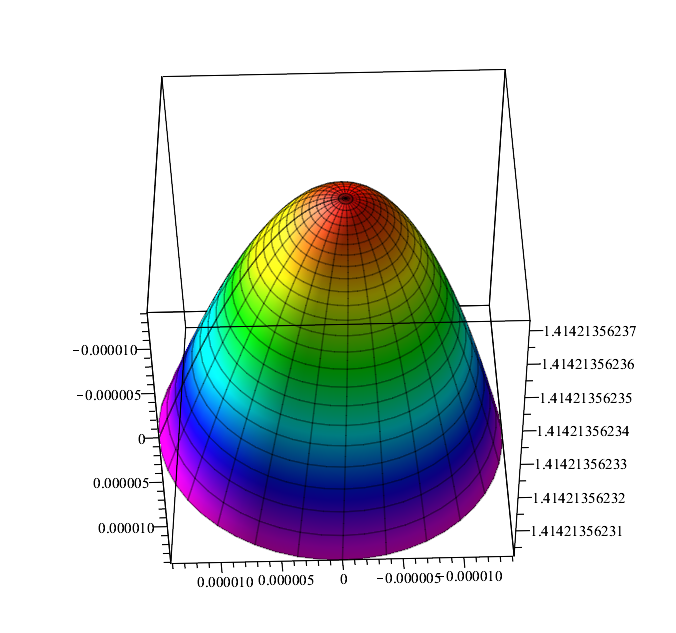} }}
\caption{The polar region of the embedding diagram (into $\mathbb{R}^3$) of a constant $r$ and constant $t$ surface for the twisted black hole. \textbf{Left:} We set $a=1$ and $r=1-0.001 < a$, and the range of the angles are $0 \leqslant \phi \leqslant  2\pi$, $0\leqslant \theta \leqslant 0.00001$. 
\textbf{Right:} Here $a=r=1$, and the angles $\phi$ and $\theta$ are plotted in the same range as the left diagram. Note that the scale of the vertical axes in both plots are either greatly stretched or compressed compared to the other two axes, otherwise the plots would appear flat (as expected; since a Riemannian manifold is locally flat).\label{embed}}
\end{figure}

Therefore, although the twisted black hole has a regular event horizon without any conical singularity, any surface $S_{a,r}$ with $a>r$ has a conical singularity at the pole with an \emph{infinite angle excess}. One way to visualize this geometry is by plotting its embedding diagram. We equate the metric \ref{2geo} with 
\begin{equation}
\d s^2 = \d z^2 + \d R^2 + R^2 \d\phi^2, 
\end{equation}
so that the surface is isometrically embedded in $\mathbb{R}^3$.
We need to solve the differential equation
\begin{equation}
\left(\frac{\d z(\theta)}{\d \theta}\right)^2 = r^2 + a^2 - \left(\frac{\d R(\theta)}{\d \theta}\right)^2,
\end{equation}
where 
\begin{equation}
R^2 (\theta) =  \frac{4a^4-4a^2r^2 - (3a^4 -6a^2r^2 -r^4)\sin^2\theta}{r^2 + a^2}.
\end{equation}
For the case $r=a$, one obtains, of course, a sphere embedded in $\mathbb{R}^3$. For $r < a$ however, the integration can only be solved numerically. Fig.(\ref{embed}) compares the region near the pole for $r<a$ (in the regime that $|r-a|$ is small) to that of $r=a$ case. We see that indeed the geometry is drastically different for the two cases. Note that despite the similarity with the Flamm paraboloid of an asymptotically flat Schwarzschild black hole (with constant $t$ and $\theta$), the left plot of Fig.(\ref{embed}) is a plot for  a constant $r$  and constant $t$ surface, and only in a small neighborhood of the north pole. \newline

It is interesting to compare this result with the so-called ``super-entropic black hole'' \cite{1504.07529}, which is a kind of asymptotically anti-de Sitter rotating black hole, whose 2-geometry of constant $r$ and constant $t$ approaches that of a hyperbolic space near the pole. Here, the interior geometry of a massless twisted black hole behaves in an ``opposite manner'' compared to the super-entropic black hole. In the latter case, the 2-surface in the embedding diagram becomes that of a ``horn'' toward the poles. See Fig.(1) of \cite{1504.07529}  for the embedding diagram of the horizon geometry. Here, however, the 2-surface in the embedding diagram ``opens wider'' toward the poles (left plot in Fig.(\ref{embed})).  Indeed, a quick calculation will reveal that $\Psi$ vanishes at the poles for the super-entropic black hole (for all values of $r$), whereas it diverges for the interior surfaces $S_{a,r}$ for the massless twisted black hole. However, the axis through the pole is actually excised from the super-entropic black hole spacetime, and the horizon is therefore non-compact. Similarly, we should remove the rotation axis from the twisted geometry, to avoid the pathology of divergent conical angle excess. \newline

One may argue that the interior of the black hole is not so important, after all, exterior observers have no access to any information behind the horizon, and so any pathological feature behind the horizon is not particularly interesting. However, for the twisted black hole, the exterior geometry is even more problematic.\newline

\subsection{Exterior Geometry}

The surfaces $S_{a,r}$ for $r>a$ cannot be globally embedded in $\mathbb{R}^3$. This in itself is not a problem. We caution the readers that one must be careful drawing any conclusion from embedding diagrams. 
Consider, for example, the Kerr-Newman
black hole. It is well-known that as the rotation becomes sufficiently fast, $J > \sqrt{3}M^2/2$, the Gaussian curvature at
the poles becomes negative, and a global isometric embedding into $\mathbb{R}^3$
is no longer possible \cite{smarr}. (A global embedding in $\mathbb{R}^4$ is achievable \cite{frolov}.) A marked difference between the twisted case and the Kerr-Newman case is that in the twisted case, for \emph{any} surface $S_{a,r}$ outside the horizon, no global embedding exists.\newline

A natural question to ask at this point is the following: what happens outside the horizon in the neighborhood of the north pole, where the ``angle excess'' becomes complex? The origin of this ``angle excess'' is that $g_{\phi\phi}$ becomes negative near the pole. Explicitly, the numerator of
\begin{equation}
 g_{\phi\phi}=\frac{4a^4-4a^2r^2 - (3a^4 -6a^2r^2 -r^4)\sin^2\theta}{r^2 + a^2}
\end{equation}
can \emph{a priori} be either positive or negative. At any fixed value of $r$, $g_{\phi\phi}$ is positive if 
\begin{equation}\label{theta}
\sin\theta >\sqrt{\frac{4a^2(r^2-a^2)}{r^4+6a^2r^2-3a^4}}.
\end{equation}\newline

This implies that for $a \neq 0$, the angle $\theta$ must be bounded away from 0 in order for $g_{\phi\phi}$ to have the ``right sign''.
One legitimate concern is that a sign change in $g_{\phi\phi}$ might result in a change of the spacetime metric signature from $(-,+,+,+)$ to $(-,-,+,+)$. We will show later that this does not happen. \newline

For now, let us look into the change of sign in $g_{\phi\phi}$, and why it remains a concern. This for three reasons.
Firstly, for Kerr black holes, any constant $t$ and constant $r$ surfaces are just ellipsoids. This agrees with the fact that the Kerr metric is a good description of the spacetime around a rotating star. If we insist that the twisted solution should in fact describe what it purported to --- two counter-rotating halves that is nevertheless static on the horizon, then ideally we would like the constant $t$ and constant $r$ surfaces to be topologically 2-sphere as well, or at least be globally Riemannian, that is, with signature $(+,+)$ throughout (away from the rotational axis). \newline

Secondly, the frame dragging angular velocity of the twisted black hole is \cite{zhang}:
\begin{equation}\label{drag}
\Omega = -\frac{g_{t\phi}}{g_{\phi\phi}}=-\frac{2a(a^2-r^2)\cos\theta}{4a^4-4a^2r^2-(3a^4-6a^2r^2-r^4)\sin^2\theta}.
\end{equation}
If $g_{\phi\phi}$ can change sign, then by continuity, it can also be zero. This means that the angular velocity will become divergent (the numerator only vanishes on the horizon, or on the equator, so it does not prevent the divergent behavior). \newline

\begin{figure}[h!]
\captionsetup{justification=raggedright}
\begin{center}
\includegraphics[width=3in]{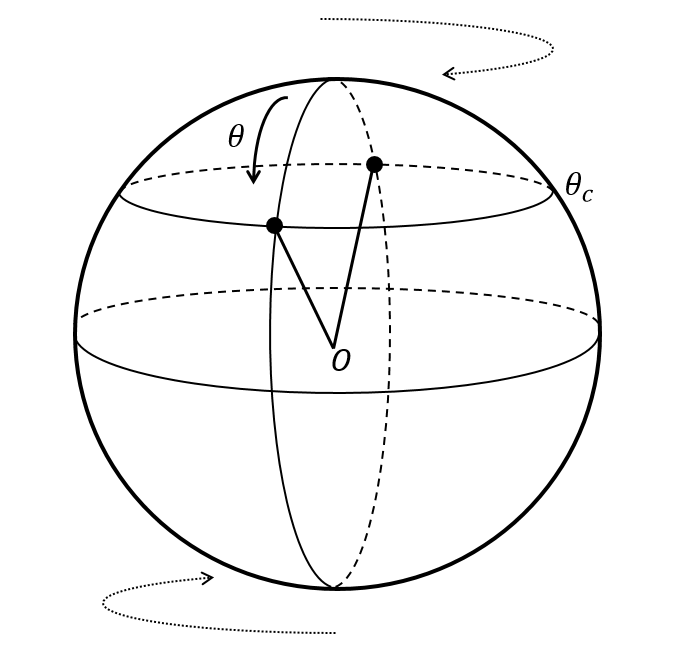} 
\caption{A surface $S_{a,r}$ for $r>a$. The dotted arrows above and below the surface indicate that the northern ``hemisphere'' is rotating in the opposite direction compared to the southern ``hemisphere''.  The point $O$ is the center of the ``sphere''.
Given a fixed $r$, the cap near the north pole below some values of $\theta$, say $\theta_C$, admits closed timelike curves ($\theta_C$ admits closed null curves).
\label{surface} }
\end{center}
\end{figure}

Thirdly,
a change of sign in $g_{\phi\phi}$ indicates that there is a closed timelike curve (CTC) in the geometry. 
The angular coordinate $\phi$ is associated to the vector field ${\partial}/{\partial \phi}$, the integral curves of which are closed (though its period might not be $2\pi$). Now 
\begin{equation}
g_{\phi \phi} = g\left(\frac{\partial}{\partial \phi},\frac{\partial}{\partial \phi}\right)
\end{equation}
is just the inner product of the vector  ${\partial}/{\partial \phi}$. Thus if this quantity becomes timelike we have a closed timelike curve. Similarly, if $g_{\phi\phi}=0$, then we have a closed null curve. See Fig.(\ref{surface}).

The situation is in fact rather similar to the well-known example of CTC spacetime: the van Stockum dust spacetime \cite{LC, VS}, whose
metric is, in the Weyl-Papapetrou form (in fact \emph{any} stationary and axisymmetric metric can be put into such a form, with suitable metric coefficients),
\begin{equation}
g_{\text{vSdust}} = -\d t^2 + 2\alpha r^2 \d t \d\phi - (1-\alpha^2r^2)r^2 \d\phi^2 - e^{-\alpha^2r^2}(\d r^2 + \d z^2), ~~\alpha > 0.
\end{equation}
The metric coefficient $g_{\phi\phi}=1- \alpha^2r^2$ vanishes when $r=1/\alpha$, where one has a closed null curve. Likewise for $r<1/\alpha$, the metric coefficient $g_{\phi\phi}$ becomes negative, and the spacetime admits CTCs. \newline

Unlike the CTCs in Kerr-Newman spacetime, which lie inside the black hole, the CTCs in the twisted spacetime occur outside the event horizon, and are therefore worrying.  The presence of CTCs also explain why the singularity theorem does not apply. \newline

If one includes the mass term, then the RHS of Eq.(\ref{theta}) becomes
\begin{equation}
\mathcal{F}(M,a,r)= \sqrt\frac{4a^2(r^2-a^2)-8M^2ar}{r^4+6a^2r^2-3a^4-8M^2ar}.
\end{equation}
However, since the numerator is only zero on the event horizon, the angle $\theta$ is always bounded away from 0, so the mass term does not prevent the problem we discussed. In fact, if the twisted black hole metric were to be used to model astrophysical black holes, then $M$ is large. For supermassive black holes, the mass term will completely dominate the expression in both the numerator and denominator at some finite coordinate distance $r$ away from the black hole, so that $\mathcal{F} \sim 1$, and so the region that admits CTCs becomes tremendously huge: the portion on the 2-geometry that is less than $\theta \sim \pi/2$ has CTCs. By symmetry, a similar region centered at the south pole also admits CTCs. Thus, only a very thin strip around the equator has no causality problem. This shows that unlike the Kerr metric, \emph{the twisted metric is not a good model of a realistic rotating black hole}.

%%%%%%%%%%%%%%%%%%%%%%%%%%%%%%%%%%%%%%%%%%%%%%%%%%%%%%%%%%%%%%%%%%%%%%%%%%%%%%%%

\section{On the Signature of the Metric Tensor}

As we mentioned previously, there is a concern that if $g_{\phi\phi}$ becomes negative, then there is a danger that the metric signature might change from $(-,+,+,+)$ to $(-,-,+,+)$. As we shall see below, this does not happen, although the metric signature does change to $(-,0,+,+)$ at the poles of any $S_{r,a}$, i.e., on the rotational axis of the black hole, which is a further evidence that the poles harbor some kind of conical singularity. \newline

Since the metric is not diagonal, it is not at all obvious what is the spacetime metric signature if $g_{\phi\phi} < 0$. One should check the eigenvalues of the full metric tensor in Eq.(\ref{twisted}). Again, the full expression is complicated and not interesting to show here. In the special case $a=1, M=0$, and $\theta=0$, however, the eigenvalues are greatly simplified:
\begin{equation}
\lambda_1 = 1+r^2, ~~\lambda_2 = \frac{1+r^2}{r^2-1}, ~~\lambda_3= \frac{\frac{5}{2}\left[(1-r^2)+|r^2-1|\right]}{r^2+1}, ~~\lambda_4= \frac{\frac{5}{2}\left[(1-r^2)-|r^2-1|\right]}{r^2+1}.
\end{equation}
Note that $\lambda_{1}$ and $\lambda_2$ are always positive outside the horizon $r_h=a=1$. However, $\lambda_3$ vanishes outside the horizon, although it is positive inside ($0 \leqslant r < 1$). On the contrary, $\lambda_4$ vanishes inside the horizon, and is negative outside.  \newline

 \begin{figure}[!h]
\captionsetup{justification=raggedright}
\centering
\includegraphics[width=3.0in]{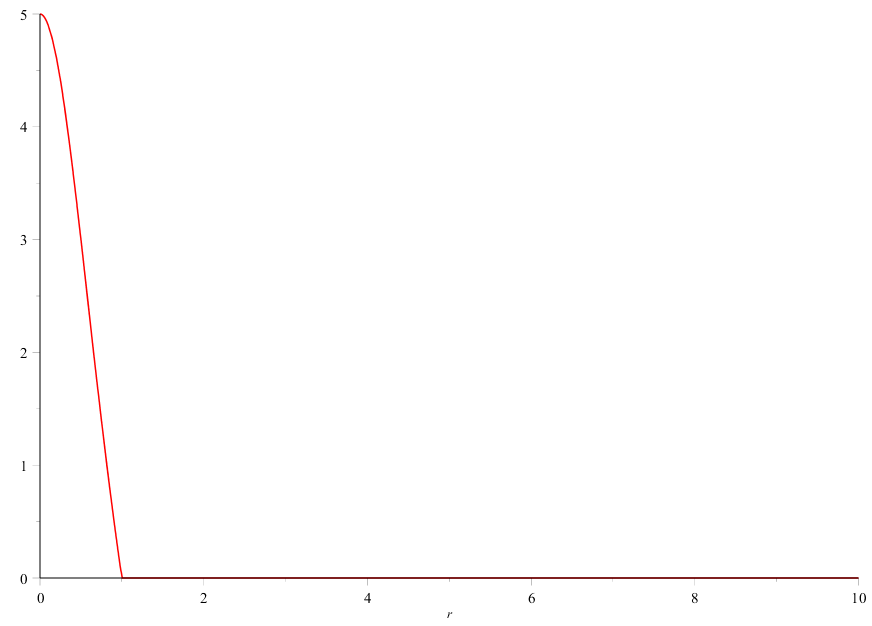}
\caption{One of the eigenvalues, which we call $\lambda_3$, of the metric tensor for a twisted black hole with $M=0, a=1$ at the north pole $\theta=0$. The eigenvalue vanishes for all $r > a$. Note that the horizontal axis starts at $r=1$, the position of the event horizon.\label{lambda}} 
\end{figure}

One might suspect that this pathological property of metric signature change is due to ``fine-tuning'' the mass term to zero. This is not the case. Restoring the mass term, for example, by setting $M=1$, $a=1$, and $\theta=0$, one finds that $\lambda_1, \lambda_2$ are as before, $\lambda_3$ is still positive inside the horizon, but zero outside. Similarly, $\lambda_4$ vanishes inside the horizon but  negative outside. Thus the metric at the poles, even in the presence of nonzero mass, has signature $(-,0,+,+)$.  
(A signature change in the metric, of the form from $(-,+,+,+)$ to $(-,+,-,-)$, can in principle also happen in the case of AdS-Kerr black hole, unless one imposes that the rotation parameter $a$ is less than the AdS curvature length scale $L$. Imposing such a condition is essential for holography to be consistent, see \cite{1504.07344}. See also \cite{1403.3258}.)\newline

In fact, to see that the mass term does not help, we can examine the determinant of the metric tensor in Eq.(\ref{twisted}). We have a surprisingly simple expression:
\begin{equation}
\text{det}(g_{\text{twist}})=-\sin^2(\theta)(a^4+2a^2r^2+r^4) \leqslant 0,
\end{equation}
which is \emph{independent of $M$}.
One easily sees that the determinant can become zero if and only if $\theta=0,\pi$. This agrees with the fact that $\lambda_3$ vanishes outside the horizon at the poles.\newline %Fig.(\ref{det}) shows a surface plot of the determinant. \newline

%\begin{figure}[h!]
%\captionsetup{justification=raggedright}
%\begin{center}
%\includegraphics[width=3.5in]{det.png} 
%\caption{The plot of the determinant of the metric tensor of the twisted spacetime, assuming that $a=1$ and $M=0$. The determinant vanishes on at the poles. \label{det}}
%\end{center}
%\end{figure}  

Therefore, we see that away from the axis of rotation the spacetime always have a Lorentzian signature $(-,+,+,+)$, which means that the spacetime is well-defined despite the fact that $g_{\phi\phi}$ can change sign. Again, let us recall that for the Kerr black hole, the constant $t$ and constant $r$ surfaces are just ellipsoids. In other words, the 3-geometry of constant $t$ is foliated by a family of ellipsoids. The fact that $g_{\phi\phi}$ can change sign means that we do not have such a foliation in the twisted case. Specifically this means that at any fixed time $t$, the twisted geometry should not be thought of as a family of (topologically) spherical surfaces indexed by $r$. However, since the spacetime remains Lorentzian, there should exist a new set of coordinates $(x^0,x^1,x^2,x^3)$ such that the family of surfaces of constant $x^0$ and constant $x^1$ are Riemannian.

%%%%%%%%%%%%%%%%%%%%%%%%%%%%%%%%%%%%%%%%%%%%%%%%%%%%%%%%%%%%%%%%%%%
\section{The Twisted Spacetime Is Taub-NUT}

It turns out that the twisted spacetime we have explored above is actually a known solution in general relativity -- it is just the Taub-NUT solution. 
The Taub-NUT metric is, in coordinates $(\mathfrak{t},r,\theta,\phi)$,
\begin{equation}
-U(r)\left[\d{\mathfrak{t}} + 4a \sin^2\left(\frac{\theta}{2}\right)\d\phi\right]^2 + \frac{\d r^2}{U(r)} + (r^2+a^2)(\d\theta^2 + \sin^2\theta d\phi^2),
\end{equation}
where 
\begin{equation}
U(r)=\frac{r^2-2Mr-a^2}{r^2+a^2}.
\end{equation}
Here we follow, except for the coordinate ${\mathfrak{t}}$, the notation in \cite{bonner}.
It is not too difficult to show that this is the same as Zhang's metric in Eq.(\ref{twisted}), 
after a simple coordinate transformation $\mathfrak{t}=t-2a\phi$. \newline

Indeed, in Eq.(\ref{twisted}),
we see that asymptotically the metric coefficient
\begin{equation}
g_{t\phi} \sim 2a \cos\theta,
\end{equation}
so that the geometry is in fact not asymptotically flat, since for asymptotically flat spacetimes, one has instead
\begin{equation}
g_{t\phi} \sim \frac{2J\sin^2\theta}{r}.
\end{equation}
In fact, the parameter $a$, which we have emphasized in the introduction that it is not a rotation parameter in the sense of Kerr black hole, is now clearly seen as the NUT charge, which has no Newtonian analog. \newline

The ``conical singularity'' at the poles that we have explored in the previous section, which corresponds to where the metric tensor becomes non-invertible, is none other than the so-called ``wire singularity'' in the Taub-NUT literature. 
Not all geodesics are complete due to these wire singularities \cite{MKG,1002.4342}.
The wire singularities can nevertheless be removed by introducing new coordinate patches and compactifying the temporal coordinate so that the topology of the spacetime becomes $\mathbb{R} \times S^3$ \cite{nut3}.
Of course, it is also known that Taub-NUT spacetime admits CTCs outside of the horizon. \newline

One obvious way to improve the geometry is to excise the regions with CTCs, and topologically identify the resulting boundaries. More specifically, after one removes the cap near the north pole (similarly for the south pole) for which $\theta \leqslant \theta_c$ (see Fig.(\ref{surface})), the leftover $S^1$ boundary can be topologically identified to a point.  This will, however, result in more conical singularities. A similar idea was proposed by Bonner back in 1969 to remove the region plagued by CTCs \cite{bonner}, but there the boundary after the excision is left as it is, thus making $S_{a,r}$ an incomplete Riemannian surface.

\section{Conclusion: The Price of Evading the Singularity Theorem}

In this work we have investigated the geometry of the 4-dimensional twisted black hole spacetime first proposed in \cite{zhang}. 
We have shown that such a twisted rotating black hole, which is free of a curvature singularity, nevertheless harbors a ``conical singularity''. This conical singularity is of a rather extreme type: inside the black hole horizon, it has an infinite amount of angle excess at the poles. On the other hand, outside the black hole horizon, the ``angle excess'' becomes complex. The Gaussian curvature is not continuous at the pole across the event horizon: it is finite on the horizon, but diverges as one approaches the horizon from either side ($-\infty$ if approaches from the inside, and $+\infty$ from the outside). Furthermore the signature of the metric along the rotational axis is $(-,0,+,+)$.\newline

We started this investigation by asking how the twisted solution evades the singularity theorem.
We now have an answer: The singularity theorem simply does not apply because the spacetime contains both closed timelike curves and a severe conical singularity along the rotational axis (except on the horizon itself).
\newline

However, none of these are new, since the twisted solution is actually the Taub-NUT metric written in a slightly different coordinate system.
The proposal to construct a rotating black hole that nevertheless has globally vanishing ADM angular momentum is an interesting one. The question is: can one construct a new black hole spacetime that is free of pathologies, including conical singularities (regardless of whether it has a curvature singularity inside the horizon), using this approach? It may well be possible that the current uniqueness theorems of black holes (see \cite{1205.6112v1} for a comprehensive review) can be extended to include rotating black hole spacetimes that have vanishing ADM angular momentum. (Note that there are other rotating black hole solutions that nevertheless have vanishing angular momentum \cite{mann2}, but these are neither solutions to pure Einstein's general relativity, nor do they achieve this via two counter-rotating halves.) \newline

\section*{Remarks} The first version of the arXiv preprint of this manuscript only pointed out that the ``twisted black hole'' spacetime contains many pathological properties. It was subsequently pointed out to the author that the spacetime is in fact just Taub-NUT. This fact was incorporated in the current, updated version, of the manuscript. On the same day version two of the manuscript appeared on the arXiv, another paper (Ref.\cite{1610.06135}) that criticizes  the twisted black hole also appeared on the arXiv.

\section*{Acknowledgement}

YCO thanks Brett McInnes, Michael Good, and Keisuke Izumi for fruitful discussions and suggestions. He thanks the following colleagues for pointing out that the ``twisted black hole''
is none other than Taub-NUT: Bahram Mashhoon, Edward Teo, Erickson Tjoa, David Chow, Cristian Stelea, G\'erard Cl\'ement, and Carlos Herdeiro. He thanks Patrick Connell for further suggestion. 
YCO also thanks Qiuyue Miao and Zhen Yuan for their much appreciated help during his recent move to Shanghai. Last but not least, YCO thanks Bin Wang and his group in the Center of Astronomy and Astrophysics for their warm welcome. He also thanks the National Natural Science Foundation of China (NNSFC) for support.

%%%%%%%%%%%%%%%%%%%%%%%%%%%%%%%%%%%%%%%%%%%%%%%%%%%%%%%%%%%%%%%%%%%%%%%%%%%%%%%%%%%%%%%%%%%%%%%%%%%%%%%%%%%%%%%%%%%%%%%%%%%%%%%%%%%%%%%%%%%%%%%%%%%%%%%%%%

%\end{multicols}


\begin{thebibliography}{99}

\bibitem{zhang}
Hongsheng Zhang, ``Twisted Spacetime in Einstein Gravity'', \href{https://arxiv.org/abs/1609.09721v1}{[arXiv:1609.09721 [gr-qc]]}.

\bibitem{1610.00886}
Songbai Chen, Jiliang Jing, ``Shadow Casted by a Twisted and Rotating Black Hole'', \href{https://arxiv.org/abs/1610.00886}{[arXiv:1610.00886 [gr-qc]]}.

\bibitem{visser}
Matt Visser, ``The Kerr Spacetime: A Brief Introduction'', \href{https://arxiv.org/abs/0706.0622v3}{[arXiv:0706.0622 [gr-qc]]}.

\bibitem{1610.06135}
Finnian Gray, Jessica Santiago, Sebastian Schuster, Matt Visser, ```Twisted' Black Holes Are Unphysical'', \href{https://arxiv.org/abs/1610.06135}{[arXiv:1610.06135 [gr-qc]]}.

\bibitem{9912320}
Richard C. Henry, ``Kretschmann Scalar for a Kerr-Newman Black Hole'', {\changeurlcolor{vividviolet}\href{http://iopscience.iop.org/article/10.1086/308819/meta;jsessionid=D72410AFF5A2AC16BC5EC51A02480F23.c2.iopscience.cld.iop.org}{Astrophys. J. \textbf{535} (2000) 350}}, \href{https://arxiv.org/abs/astro-ph/9912320}{[arXiv:astro-ph/9912320]}.

\bibitem{penrose}
 Roger Penrose, ``Gravitational Collapse and Space-Time Singularities'', {\changeurlcolor{vividviolet}\href{https://journals.aps.org/prl/abstract/10.1103/PhysRevLett.14.57}{Phys. Rev. Lett. \textbf{14} (1965) 57}}.


\bibitem{jose}
Jos\'e M. M. Senovilla, David Garfinkle, ``The 1965 Penrose Singularity Theorem'', {\changeurlcolor{vividviolet}\href{http://iopscience.iop.org/article/10.1088/0264-9381/32/12/124008/meta;jsessionid=64FEF6A139E0DDDE26DC1F3A4488EEE5.c1.iopscience.cld.iop.org}{Class. Quant. Grav. \textbf{32} (2015) 124008}}, \href{https://arxiv.org/abs/1410.5226}{[arXiv:1410.5226 [gr-qc]]}.

\bibitem{parikh}
Maulik Parikh, ``Two Roads to the Null Energy Condition'',  {\changeurlcolor{vividviolet}\href{https://arxiv.org/abs/1512.03448}{Int. J. Mod. Phys. D \textbf{24} (2015) 1544030}}, \href{https://arxiv.org/abs/1512.03448}{[arXiv:1512.03448 [hep-th]]}.


\bibitem{1502.00287}
Michael Kunzinger, Roland Steinbauer, James A. Vickers, ``The Penrose Singularity Theorem in Regularity $C^{1,1}$'', {\changeurlcolor{vividviolet}\href{https://iopscience.iop.org/0264-9381/32/15/155010/}{Class. Quant. Grav. \textbf32 (2015) 155010}}, \href{https://arxiv.org/abs/1502.00287}{[arXiv:1502.00287 [gr-qc]]}.

\bibitem{kinnersley}
William Kinnersley, ``Type D Vacuum Metrics'', {\changeurlcolor{vividviolet}\href{http://scitation.aip.org/content/aip/journal/jmp/10/7/10.1063/1.1664958}{J. Math. Phys. \textbf{10} (1969) 1195}}.

\bibitem{nut1}
Abraham H. Taub,  ``Empty Space-Times Admitting a Three Parameter Group of Motions'', {\changeurlcolor{vividviolet}\href{https://www.jstor.org/stable/1969567?origin=crossref&seq=1#page_scan_tab_contents}{Annals of Math. \textbf{53} (1951) 472}}.

\bibitem{nut2}
Ezra T. Newman,  L. Tamburino, T. Unti, ``Empty-Space Generalization of the Schwarzschild Metric'', {\changeurlcolor{vividviolet}\href{http://scitation.aip.org/content/aip/journal/jmp/4/7/10.1063/1.1704018}{J. Math. Phys., \textbf{4} (1963) 91}}.

\bibitem{nut3}
Charles W. Misner, ``The Flatter Regions of Newman, Unti, and Tamburino's Generalized Schwarzschild Space'', {\changeurlcolor{vividviolet}\href{http://scitation.aip.org/content/aip/journal/jmp/4/7/10.1063/1.1704019}{J. Math. Phys. \textbf{4} (1963) 924}}.

\bibitem{MKG}
J. G. Miller, Martin D. Kruskal, and Brendan B. Godfrey, ``Taub-NUT (Newman, Unti, Tamburino) Metric and Incompatible Extensions'', {\changeurlcolor{vividviolet}\href{https://journals.aps.org/prd/abstract/10.1103/PhysRevD.4.2945}{Phys. Rev. D \textbf{4} (1971) 2945}}.

\bibitem{1002.4342}
Valeria Kagramanova, Jutta Kunz, Eva Hackmann, Claus Laemmerzahl, ``Analytic Treatment of Complete and Incomplete Geodesics in Taub-NUT Space-Times'', {\changeurlcolor{vividviolet}\href{https://arxiv.org/abs/1002.4342v2}{Phys. Rev. D \textbf{81} (2010) 124044}}, \href{https://arxiv.org/abs/1002.4342v2}{[arXiv:1002.4342 [gr-qc]]}.

\bibitem{1508.07622}
G\'erard Cl\'ement, Dmitri Gal'tsov, Mourad Guenouche, ``Rehabilitating Space-Times with NUTs'', {\changeurlcolor{vividviolet}\href{http://www.sciencedirect.com/science/article/pii/S0370269315007492}{Phys. Lett B. \textbf{750} (2015) 591}}, \href{https://arxiv.org/abs/1508.07622}{[arXiv:1508.07622 [hep-th]]}.

\bibitem{1504.07529}
Robie A. Hennigar, David Kubiznak, Robert B. Mann, Nathan Musoke, ``Ultraspinning Limits and Super-Entropic Black Holes'', {\changeurlcolor{vividviolet}\href{https://link.springer.com/article/10.1007\%2FJHEP06\%282015\%29096}{JHEP \textbf{1506} (2015) 096}}, \href{https://arxiv.org/abs/1504.07529}{[arXiv:1504.07529 [hep-th]]}.

\bibitem{smarr}
Larry Smarr, ``Surface Geometry of Charged Rotating Black Holes'', {\changeurlcolor{vividviolet}\href{https://journals.aps.org/prd/abstract/10.1103/PhysRevD.7.289}{Phys. Rev. D \textbf{7} (1973) 289}}.

\bibitem{frolov}
Valeri P. Frolov, ``Embedding of the Kerr-Newman Black Hole Surface in Euclidean Space'', {\changeurlcolor{vividviolet}\changeurlcolor{vividviolet}\href{https://journals.aps.org/prd/abstract/10.1103/PhysRevD.73.064021}{	Phys. Rev. D \textbf{73} (2006) 064021}}, \href{https://arxiv.org/abs/gr-qc/0601104v1}{[arXiv:gr-qc/0601104]}.



\bibitem{LC}
Cornelius Lanczos, ``\"Uber eine station\"are Kosmologie im Sinne der Einsteinschen Gravitationstheorie'', {\changeurlcolor{vividviolet}\href{https://link.springer.com/article/10.1007\%2FBF01328251}{Zeitschrift f\"ur Physik \textbf{21} (1924) 73}}.

\bibitem{VS}
Willem J. van Stockum, ``The Gravitational Field of a Distribution of Particles Rotating Around an Axis of Symmetry'', {\changeurlcolor{vividviolet}\href{https://www.cambridge.org/core/journals/proceedings-of-the-royal-society-of-edinburgh/article/ixthe-gravitational-field-of-a-distribution-of-particles-rotating-about-an-axis-of-symmetry/40E39372658C7031B0C4316A36154F46}{Proc. Roy. Soc. Edinburgh A. \textbf{57}  (1937) 135}}.



\bibitem{1504.07344}
Brett McInnes, Yen Chin Ong, ``When Is Holography Consistent?'', {\changeurlcolor{vividviolet}\href{http://www.sciencedirect.com/science/article/pii/S0550321315002412}{Nucl. Phys. B \textbf{898} (2015) 197}}, \href{https://arxiv.org/abs/1504.07344}{[arXiv:1504.07344 [hep-th]]}.

\bibitem{1403.3258}
Brett McInnes, ``Angular Momentum in QGP Holography'', {\changeurlcolor{vividviolet}\href{http://www.sciencedirect.com/science/article/pii/S0550321314002661}{Nucl. Phys. B \textbf{887} (2014) 246}}, \href{https://arxiv.org/abs/1403.3258}{[arXiv:1403.3258 [hep-th]]}.

\bibitem{bonner}
William B. Bonnor, ``A New Interpretation of the NUT Metric in General Relativity'', {\href{https://www.cambridge.org/core/journals/mathematical-proceedings-of-the-cambridge-philosophical-society/article/a-new-interpretation-of-the-nut-metric-in-general-relativity/3C7F5F3E0DC2F1B355F2B0D195EEE0D1}{Proc. Camb. Phil. Soc. \textbf{66} (1969) 145}}.


\bibitem{1205.6112v1}
Piotr T. Chru\'sciel, Jo$\tilde{\text{a}}$o Lopes Costa, Markus Heusler, ``Stationary Black Holes: Uniqueness and Beyond'',  {\changeurlcolor{vividviolet}\href{http://relativity.livingreviews.org/Articles/lrr-2012-7/}{Living Rev. Relativity \textbf{15} (2012) 7}}, \href{https://arxiv.org/abs/1205.6112v1}{[arXiv:1205.6112 [gr-qc]]}.

\bibitem{mann2}
Mozhgan Mir, Robert B. Mann, ``Charged Rotating AdS Black Holes with Chern-Simons Coupling'', {\changeurlcolor{vividviolet}\href{https://journals.aps.org/prd/abstract/10.1103/PhysRevD.95.024005}{Phys. Rev. D \textbf{95} (2017)  024005}}, \href{https://arxiv.org/abs/1610.05281}{[arXiv:1610.05281 [gr-qc]]}.









\end{thebibliography}
\end{document}